\documentclass[prl,twocolumn]{revtex4}
\usepackage{amsmath}
\usepackage{dcolumn}
\usepackage{bm}
\usepackage{epsfig}
\usepackage{epstopdf}
\usepackage{graphicx}
\usepackage{amsfonts}
\usepackage{amssymb}
\usepackage{mathrsfs}
\usepackage{color}
\usepackage{multirow}
\usepackage{appendix}

\begin{document}

\title{Energy-conversion device using a quantum engine with the work medium of two-atom entanglement}
\author{J.-W. Zhang$^{1}$} \thanks{Co-first authors with equal contribution}
\author{B. Wang$^{2,3}$}  \thanks{Co-first authors with equal contribution}
\author{W.-F. Yuan$^{2,3}$} \thanks{Co-first authors with equal contribution}
\author{J.-C. Li$^{2,3}$} \thanks{Co-first authors with equal contribution}
\author{J.-T. Bu$^{2,3}$}
\author{G.-Y. Ding$^{2,3}$}
\author{W.-Q. Ding$^{2,3}$}
\author{L. Chen$^{2,1}$}
\author{F. Zhou$^{2,1}$}
\email{zhoufei@wipm.ac.cn}
\author{M. Feng$^{1,2,4}$}
\email{mangfeng@wipm.ac.cn}
\affiliation{$^{1}$Research Center for Quantum Precision Measurement, Guangzhou Institute of Industry Technology, Guangzhou, 511458, China \\
	$^{2}$State Key Laboratory of Magnetic Resonance and Atomic and Molecular Physics,
	Wuhan Institute of Physics and Mathematics, Innovation Academy of Precision Measurement Science and Technology, Chinese Academy of Sciences, Wuhan, 430071, China\\
	$^{3}$School of Physics, University of the Chinese Academy of Sciences, Beijing 100049, China \\
	$^{4}$Department of Physics, Zhejiang Normal University, Jinhua 321004, China }


\begin{abstract}
Although entanglement is considered as an essential resource for quantum information processing, whether entanglement helps for energy conversion or output in the quantum regime is still lack of experimental witness. Here we report on an energy-conversion device operating as a quantum engine with the working medium acted by two entangled ions confined in a harmonic potential. The two ions are entangled by virtually coupling to one of the vibrational modes shared by the two ions, and the quantum engine couples to a quantum load, which is another shared vibrational mode. We explore the energy conversion efficiency of the quantum engine and investigate the useful energy (i.e., the maximum extractable work) stored in the quantum load by tuning the two ions in different degrees of entanglement as well as detecting the change of the phonons in the load.
Our observation provides, for the first time, quantitative evidence that entanglement fuels the useful energy produced by the quantum engine, but not helpful for the energy conversion efficiency. We consider that our results may be useful to the study of quantum batteries for which one of the most indexes is the maximum extractable energy.  \\
\\
\end{abstract}
\maketitle
Quantum engines (QEs) operated with quantum substances are expected to surpass their classical counterparts in terms of output power and efficiency by utilizing quantum features~\cite{Scully,Gemmer,Parrondo,Chapin,Robsnagel1,Hardal}. The recently growing interest in QEs is additionally motivated by the experiments of single-spin QEs executed from trapped-ion systems to optomechanics \cite{Klatzow,Robsnagel2,Maslennikov,Lindenfels,Zhang2,Chiral,Quan,Guthrie,Ryan,Peterson,Koski,Zhang,Dechant}.

Entanglement~\cite{Schrodinger} is a unique resource in quantum information processing, which speeds up computation~\cite{Kendon}, ensures the information security in communication~\cite{Ekert,Bennett1,Bennett2} and improves the signal-to-noise ratio in precision measurement~\cite{Venzl,Wineland,Huelga,Giovannetti,Jozsa}. Recently, various ideas with entanglement involved in QEs have been proposed ~\cite{Hewgill,Barrios,Tavakoli,Josefsson,Huang,Bresque}, indicating that entanglement helps improve the efficiency of the engine over the classical counterparts. Experimentally, enhanced performance of the  QE was demonstrated with linear optics \cite{Wang} due to entanglement between different degrees of freedom in single photons as well as local measurements. However, a deeper understanding of the QE performance associated with entanglement needs quantitatively tuning entanglement in the working medium, which is hard for single-photon experiments.

In the present work, by quantitatively varying the degree of entanglement, we experimentally investigate, for the first time, the energy conversion from the QE involving bipartite entanglement to a quantum load \cite{explain}, in which the useful energy extracted (also referred to as the work extraction) from the quantum load is focused. As clarified later, we witness that the entangled working medium of QE can fuel the work extraction from the load, but not helpful for the efficiency of the energy conversion from QE to the load.
\begin{figure*}[htbp]
	\centering
	\includegraphics[width=18.1 cm]{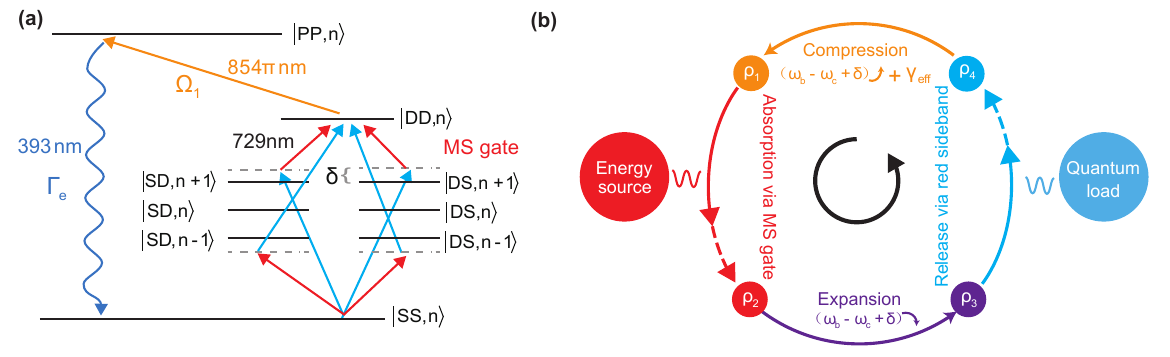}
	\caption{(a) Level scheme of two $^{40}$Ca$^+$ ions entangled via the breath mode state $|n\rangle$ of the axial vibration. The arrows in red and blue represent, respectively, the red- and blue-detuning irradiation of the 729-nm laser for achieving the MS gate, where $\delta$ is the value of the symmetric detuning. The $\pi$-polarized 854-nm laser beam and the 393-nm spontaneous emission construct the dissipative channel. (b) A quantum thermodynamic cycle, where the first stroke absorbs photons by implementing the MS gate; The second stroke operates with declining detuning; The third stroke is carried out by red-sideband transition to couple to the load (i.e., the center-of-mass mode), and the fourth stroke encloses the cycle by enlarging the detuning under dissipation.}
	\label{Fig1}
\end{figure*}

The working medium of our QE is composed of two $^{40}$Ca$^{+}$ ions confined in a linear Paul trap. We encode the qubits into the pseudo-spin states of each of the ions, i.e., $|S\rangle = |4^2S_{1/2}, m_J = -1/2 \rangle$ and $|D\rangle = |3^2D_{5/2}, m_J = -3/2 \rangle$ with the magnetic quantum number $m_J$. The two-ion system owns two vibrational modes, i.e. the center-of-mass mode (with frequency of $\omega_c/2\pi = 0.794$ MHz) and the breath mode (with frequency of $\omega_b/2\pi = 1.37$ MHz) along the axial direction.
To reduce the detrimental effect from decoherence, we employ the breath mode to help entangle the two ions by the M$\phi$lmer-S$\phi$rensen (MS) gate \cite{MSgate}, in which the two qubits are virtually coupled to the breath mode under global irradiation of 729-nm laser beams.
Besides, we detect the output energy by measuring the variation of the average phonon number in the load (i.e., the axial center-of-mass mode). 

The quantization axis is parallel to the axial direction by a magnetic field applied with approximately 3.45 Gauss at the trap center. Prior to the experiment, we have accomplished both the Doppler and sideband cooling, which reduces the thermal phonons to the average phonon number $n_{\rm h} \approx$  0.03 of the breath mode and $n_{\rm c} \approx$ 0.13 of the center-of-mass vibrational mode with the Lamb-Dicke parameter of $\eta_p = 0.0574$. This is sufficient to avoid thermal phonons yielding offsets of Rabi oscillations, and ensure correct observation of the output energy from the phonons. Figure~{\ref{Fig1}}(a) sketches the entanglement realization with the MS gate, where the two ions are driven globally by a red-detuned and a blue-detuned 729-nm laser beams with the symmetric detuning $\delta$ from the breathe mode. By exactly tuning the duration of the 729-nm laser irradiation, we may achieve, from the initial state $|SS\rangle$, entanglement between $|SS\rangle$ and $|DD\rangle$.
\begin{figure*}[htbp]
	\centering
	\includegraphics[width=17.5 cm]{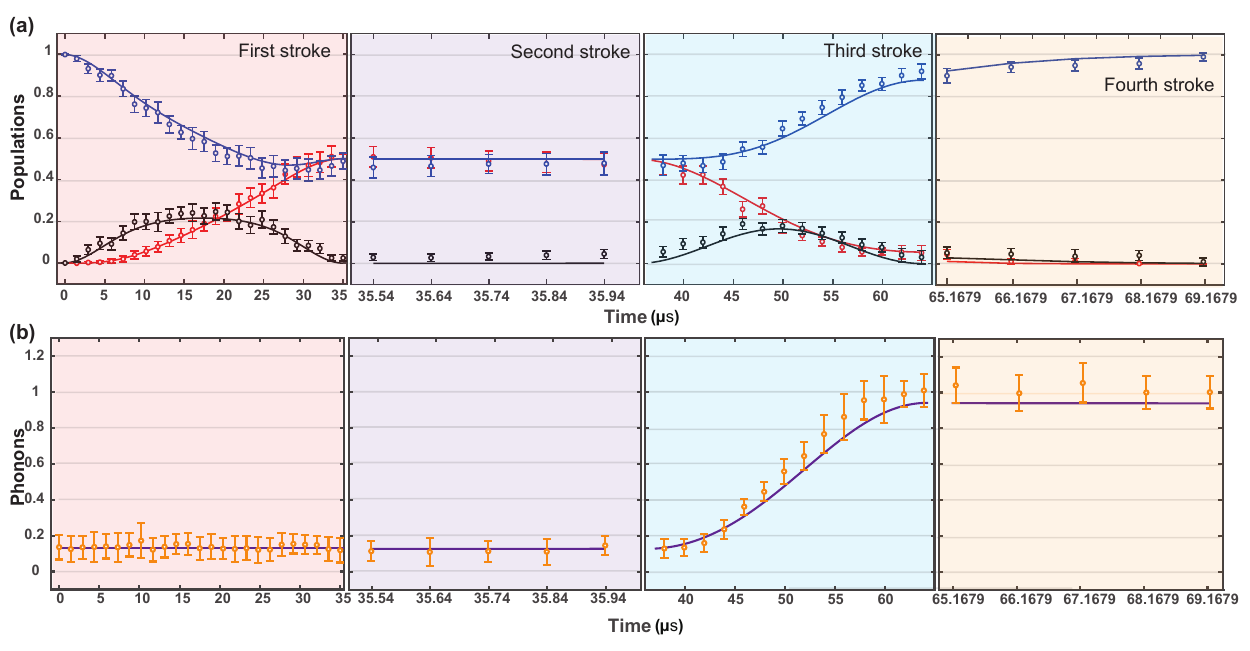}
	\caption{Experimental measurement of energy conversion. (a) Time evolution of the populations, where $|SS\rangle$, $|SD\rangle +|DS\rangle$, and $|DD\rangle$, are colored, respectively, in blue, black, and red. (b) Time evolution of the mean number of the phonons. The lines are obtained by theoretical simulation and dots are experimental results. The error bars indicating the statistical standard deviation of the experimental data are obtained by 10 000 measurements for each data point.}
	\label{Fig2}
\end{figure*}

For our purpose (e.g., the fourth stroke as plotted in Fig.~{\ref{Fig1}}(b)), we must introduce a dissipative channel for the qubits. To this end, we employ an extra energy level $|4^2P_{3/2}, m_J = -3/2 \rangle$ (labeled as $|P\rangle$), which couples to $|D\rangle$ by a 854-nm laser (with Rabi frequency $\Omega_1$ and $\pi$ polarization) and dissipates to $|S\rangle$ by spontaneous emission with the decay rate of $\Gamma_e/2\pi = 23.1$ MHz (see Fig.~{\ref{Fig1}}(a)). This achieves, under appropriate laser irradiation, an effective two-level model with the decay $\gamma_{\rm eff} = \Omega^2_1/\Gamma_e$~\cite{Zhang1,Zhang2}, and described by the Lindblad master equation as
$\dot{\rho}_{m} =-i[H_s, \rho_{m}] + \sum\limits_{i=1}^{2}\frac{\gamma_{\rm eff}}{2}(2\sigma_{-}^i\rho_{m}\sigma_{+}^i -\sigma_{+}^i\sigma_{-}^i\rho_{m} - \rho_{m}\sigma_{+}^i\sigma_{-}^i)$,
where $i$ is the denotation representing the $ith$ ion, $\rho_{m}$ denotes the density operator of the working medium, $H_s$ is the Hamiltonian of the working medium in the fourth stroke, and $\sigma_{-} = (\sigma_{+})^{\dagger}\equiv |g\rangle\langle e|$ is the annihilation operator of the qubit.

The operations of this QE are depicted in Fig.~{\ref{Fig1}}(b), which consists of four strokes. The first stroke is the heating stroke for absorbing energy, which is operated by the MS gate and governed by the Hamiltonian
\begin{equation}
\begin{split}
	H_s^{\rm 1st} &= \sum\limits_{i=1}^{2}\frac{\eta_p\Omega_r}{2}({\sigma_{+}^ie^{\rm i\delta_rt}a_h+\sigma_{-}^ie^{\rm-i\delta_rt}a^{\dagger}_h}) \\
	 &~~+ \sum\limits_{i=1}^{2}\frac{\eta_p\Omega_b}{2}({\sigma_{+}^ie^{\rm i\delta_bt}a^{\dagger}_h+\sigma_{-}^ie^{\rm -i\delta_bt}a_h}),
	 \label{Eq2}
\end{split}
\end{equation}
where $\Omega_r$ ($\Omega_b$) and $\delta_r$ ($\delta_b$) represent the Rabi frequency and detuning of the red (blue) sideband transition of the breath vibration mode, respectively, $a_h$ and $a^{\dagger}_h$ are phononic operators associated with the axial breath vibration mode.

In our experiment, we may simply consider $\delta_r =-\delta_b = -\delta$, $\Omega_r = \Omega_b =\Omega$. Throughout this work, we fix $\Omega$ to be $\Omega/2\pi = 245.8$ kHz.
In this first stroke, we produce different degrees of entanglement of the two ions by exactly tuning the MS-gating time, which corresponds to the absorption of different numbers of photons, as clarified below. Experimentally, the degree of entanglement produced by the MS gate is quantized by the gating fidelity \cite{SM}. For the working medium initialized from $|SS\rangle$, the absorbed optical photon quanta $\Delta n_t$ is defined by $2P_{\rm DD}+P_{\rm SD}+P_{\rm DS}$ \cite{Horne}, where $P_{\rm DD}$, $P_{\rm SD}$ and $P_{\rm DS}$ represent the populations of $|DD\rangle$, $|SD\rangle$ and $|DS\rangle$, respectively.

Then we rapidly tune the laser frequency to couple the center-of-mass vibrational model.
In this case, the two ions interact with the globally irradiated 729-nm laser respectively, given by the Hamiltonian $H_s^{\rm 2nd}=\sum\limits_{i=1}^{2}\frac{\eta_p\Omega}{2}\sigma_{+}^i(e^{-\rm i\int^{0}_{\omega_b-\omega_c+\delta}\delta(t)dt}a_c+e^{\rm i\int^{0}_{\omega_b-\omega_c+\delta}\delta(t)dt}a^{\dagger}_c) + H.c.$ with $a_c$ and $a^{\dagger}_c$ representing phononic operators associated with the axial center-of-mass mode. Due to the rapid change of laser frequency in a very short duration, the populations in both the qubit states and the vibrational mode remain unchanged during this stroke. The third stroke aims to transform the internal energy (represented by the number of the absorbed optical photons) to the external energy (represented by the number of the produced phonons in the load), which is accomplished by a Jaynes-Cummings interaction under the government of the Hamiltonian $H_s^{\rm 3rd}=\sum\limits_{i=1}^{2}\frac{\eta_p\Omega}{2}(\sigma_{+}^{i}a_c + \sigma_{-}^{i}a^{\dagger}_c)$.
The fourth stroke is the time-reversed process of the second stroke. However, in this stroke, for enclosing the cycle, we switch on the dissipative channels in the two ions, i.e., $\gamma_{\rm eff}\neq$ 0.
\\
\begin{figure*}[htbp]
	\centering
	\includegraphics[width=17.5 cm]{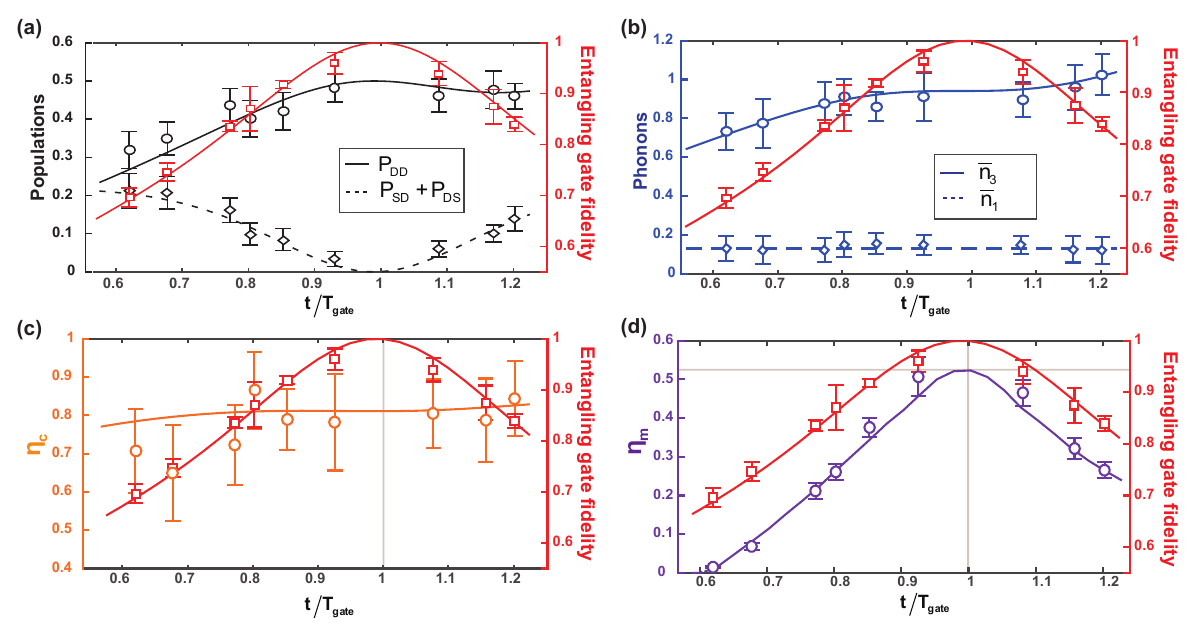}
	\caption{Time evolution of the characteristic parameters of the system. (a) Time evolution of populations and entanglement of the qubits. (b) Time evolution of the phonons as well as  entanglement  of the qubits, where the dashed line $\bar{n}_1$ and the solid line $\bar{n}_3$ represent, respectively, the mean phonon numbers in the final states of the first and third strokes. (c,~d) Time evolution of the conversion efficiency $\eta_c$ and mechanical efficiency $\eta_m$ in comparison with entanglement variation of the qubits. The vertical gray line represents the time point for maximum entanglement. In all the panels, $T_{\rm gate}= 36$ $\mu$s. Lines are obtained by simulation and dots are experimental results. The error bars indicating the statistical standard deviation of the experimental data are obtained by 10 000 measurements for each data point. The relatively large error bars presented in (c,~d) are due to the fact that detecting phonon numbers with ultimate precision is intricate using current techniques.}
	\label{Fig3}
\end{figure*}

Before specifying the experimental details, we elucidate some key points relevant to the energies and efficiencies of the QE under our consideration.
We evaluate the number of the absorbed optical photons by the net increase of populations in the qubit levels \cite{Horne}, and the converted energy to the load by the the number of the produced phonons in the center-of-mass mode. Moreover, since the energy of the phonons is tiny compared with that of the photons, it is not helpful to define the QE efficiency in terms of energies. To evaluate the effect of the energy conversion from optical photons to vibrational phonons, we follow the idea in \cite{Horne} by considering the efficiency of conversion of quanta, i.e., by defining the conversion efficiency of our QE as $\eta_c = \Delta n_{t}/\Delta n_{o}$, where $\Delta n_t$ and $\Delta n_o$ represent the net increase of the mean phonon number in the load and the absorbed photon quanta in the working medium, respectively. Furthermore, for quantitatively figuring out the useful energy extracted from output phonons as well as the associated efficiency, we employ the concepts of ergotropy $\mathcal{W}$ \cite{ergotropy} and quantum mechanical efficiency $\eta_m = \mathcal{W}/(\Delta n_o\hbar\omega_{c})$ where $\Delta n_o\hbar\omega_{c}$ represents the obtained energy of the output phonons. Ergotropy is defined by the maximum energy extracted using unitary operations, quantified by the difference in energy between the original state $\rho_p$ and the passified state $\widetilde{\rho}_p$, i.e., $\mathcal{W} = tr[H_p\rho_p]- tr[H_p\widetilde{\rho}_p]$, where $H_p$ is the Hamiltonian of the phonons, the passified state corresponds to the considered density matrix being diagonal in the representation of energy eigenstates with the diagonal terms (i.e., the occupation probabilities) decreasing for the enlarging energy eigenvalues~\cite{passive1,passive2,Francica}.


Experimentally, we first execute the QE with the maximum entanglement produced in the first stroke, i.e., with the fidelity $F=0.9625$ with respect to the perfect case \cite{SM}. This is accomplished by the MS gate for a time duration $T_{\rm gate} = 2\pi/\delta \approx 36$ $\mu$s with $\delta = 2 \eta_p \Omega $. From Figs.~{\ref{Fig2}}(a,b), we observe the time evolution of both the populations in qubit states and the mean number of phonons, respectively. At the end of this stroke, we have $P_{\rm DD} \approx 0.4851$ and $P_{\rm SD} + P_{\rm DS}\approx 0.02541$, indicating the absorbed optical photons $\Delta n_o = 0.99561$\cite{SM}. In the second stroke, rapidly reducing the detuning of the center-of-mass mode from 859 kHz to 0 is accomplished by four successive steps with each of 0.1 $\mu$s. As monitored in Fig.~{\ref{Fig2}}(a,b), the populations in qubit states and the mean number of phonons are almost unchanged during this process which is of very short time duration.
The third stroke is very crucial in our QE cycle, which outputs the energy of the QE to the load. This is achieved by a red-sideband transition of the 729-nm laser irradiating globally the two ions.
After well compensating the Stark effect, we have found the optimal duration of the laser irradiation for the energy conversion, which is 26.5 $\mu$s, leading to $\Delta n_t \approx 0.91-0.13 = 0.78$ (for excluding the initial value 0.13).
In this stroke, to experimentally evaluate the ergotropy of the output phonons, we measure the populations $P_n$, which are the diagonal elements of the phononic density matrix $\rho_p$, by fitting the experimental data of the blue-sideband transition of the two ions \cite{SM}.
This is an approximate treatment based on the assumption that the off-diagonal elements in $\rho_p$ are negligible \cite{Horne}.
Although it is not as good as in the single-qubit case \cite{Horne}, this approximate treatment has the nearly identical variation to the exact result \cite{SM}.  Therefore, due to the initial thermal state (passified state) of the load, we estimate approximately the useful energy of the load to be $\mathcal{W}/\hbar\omega_{c} \approx 0.373$. The fourth stroke enlarges the detuning to the center-of-mass mode from 0 to 859 kHz using five successive steps with each of 1 $\mu$s. Since this is the last stroke, we have to enclose the QE cycle by switching on the dissipative channel. We turn on the 854-nm laser, introducing an effective decay with $\gamma_{\rm eff} = 5\Omega$.
Therefore, the efficiencies of this QE are  $\eta_c=0.7834$ and  $\eta_m = 0.49$, respectively.

To have deeper insight into the role of entanglement in the QE, we need to quantitatively evaluate the performance of the QEs by tuning the degree of entanglement of the working medium in the first stroke. This is accomplished by exactly tuning the gating time $t$ due to the fixed value of $\Omega$. Experimentally, we choose multiple time intervals around the $T_{\rm gate}$ \cite{SM} and monitor variation of the absorbed optical quanta as well as the mean numbers of the net increased phonons, as illustrated in Fig.~{\ref{Fig3}}(a,~b). From these values, we acquire the efficiency $\eta_c$ at different degrees of entanglement, see Fig.~{\ref{Fig3}}(c). We also investigate the useful energy stored in the load by fitting the experimental data of the blue-sideband transition at the end of third stroke \cite{SM}, see Fig.~{\ref{Fig3}}(d). It is evident that
the maximum value of $\eta_m$ appears at the maximum entanglement of the working medium with values of $\mathcal{W}/\hbar\omega_{c} \approx 0.4242$, $\Delta n_{t}=0.8108$ and $\eta_{m}=0.523$ \cite{SM}, while $\eta_c$ remains nearly constant with respect to the change of entanglement. This indicates that the working medium in entanglement is not helpful for energy conversion, but strongly supports the mechanical efficiency. Considering the exact solution of ergotropy that varies from zero to the maximum with the variation of entanglement, we conjecture that, it is the entanglement in our case that fuels the useful energy in the load \cite{SM}.


In summary, we have experimentally executed a QE with the working medium of tunable bipartite entanglement, working as an energy-conversion device to output energy to a quantum load. The impact of entanglement of the working medium on the quantum load has been witnessed quantitatively by measuring the efficiencies of $\eta_c$ and $\eta_m$. In particular, the study of the ergotropy and $\eta_m$ associated with the maximum extractable energy of the load would be helpful for understanding the process of charging the load, if the vibrational mode of the ions is acted as a quantum battery \cite{Yang}. Therefore, our findings would inspire further investigation of microscopic energy devices, e.g., QEs and quantum batteries, with entanglement involved.

This work was supported by National Natural Science Foundation of China under Grant Nos. 12304315, U21A20434, 12074346, 12074390, by China Postdoctoral Science Foundation under Grant Nos. 2022M710881, 2023T160144, by Key Lab of Guangzhou for Quantum Precision Measurement under Grant No. 202201000010, by Science and Technology Projects in Guangzhou under Grant Nos. 202201011727 and 2023A04J0050, and by Nansha Senior Leading Talent Team Technology Project under Grant No. 2021CXTD02.
\\


\begin{thebibliography}{99}
\bibitem{explain} In the execution of the QE cycles, the coupling of the working medium and the quantum load develops correlations between them and thus increases information entropy in the load. So here it is impossible to clearly define heat or work transfer as usual. We prefer to call it as energy transfer from the QE to the load.
\end{thebibliography}
\end{document}